\documentclass[12pt]{article}
\usepackage{amssymb}
\usepackage{amsmath}
\usepackage{epsfig}
\newcommand{\be}{\begin{equation}}
\newcommand{\ee}{\end{equation}}
\newcommand{\bea}{\begin{eqnarray}}
\newcommand{\eea}{\end{eqnarray}}
\DeclareMathOperator {\Res}{Res}

\makeatother
\makeatletter
\renewcommand{\@biblabel}[1]{#1.}
\makeatother
\date{}

\begin{document}

\begin{center}
{\large\bf A proposed solution for the lifetime puzzle of the $^{229m}$Th$^+$ isomer
}

\large
\bigskip
{ F. F. Karpeshin$^1$, M. B. Trzhaskovskaya$^2$ }\\
      $^1$D. I. Mendeleyev Institute for Metrology, Saint-Petersburg, Russia

     $^2$National Research Center``Kurchatov Institute'' --- Petersburg Nuclear Physics
      Institute, Russia

 \vspace{0.1cm}
{\it E-mail: fkarpeshin@gmail.com}
\\
\end{center}

\smallskip \smallskip \bigskip  \normalsize \noindent
With the example of the $^{229}$Th nucleus, which is the most
likely candidate for the creation frequency standard of a future,
the dynamics of the interplay and the relationship of various
resonance conversion mechanisms is analyzed. As a result, a
solution is proposed for  the so-called thorium puzzle, which
consisted of a  contradiction between  the experimental and
theoretical  lifetimes of Th$^+$ ions. First, the solution
demonstrates  the dependence of the lifetime of the nuclear isomer
on the ambient conditions. Second, it demonstrates the leveling
role of the fragmentation of the single-electron levels, which
makes the resonance amplification of the electron-nuclear
interaction more likely. Both of these trends lead to a probable
decrease of the theoretical lifetime towards agreement with
experiment.

\large
\bigskip

\begin{center}
%{''…"…ˆ…}
\end{center}

   \section{Introduction}

$^{229}$Th is a nucleus with a uniquely narrow doublet of the ground and isomeric first excited level, split by only $\sim$8 eV (\cite{beck1,beck2,seif2,yama,geist2} and Refs. cited therein). For this reason, its observation and measurement was a longstanding problem. Whereas the first suggestions of its existence appeared decades ago \cite{reich}, it was only in \cite{lars} that its conversion decay was directly detected for the first time.  The measured lifetime of about 10 $\mu$s in neutral atoms \cite{seif} precisely coincided with the predicted value \cite{ka07}. The isomer undergoes  deexcitation by a radiative $M1$ transition, with its own half-life of  about 2 h, which corresponds to a radiative width of  $\sim10^{-20}$ eV.  Two years after the first observation, the reality of the isomer was confirmed by direct detection of its hyperfine structure \cite{ptb}.

   $^{229}$Th has the lowest isomer energy amongst the known nuclei. It is the most likely candidate for the frequency reference and nuclear-optical clock of the next generation \cite{peik}. For a long time, its energy was considered to be even less, about 3.5 eV. At such an energy, the nuclear level is entangled with many electronic levels with the same multipolarity. This means that we never can observe photons exclusively originating from its deexcitation.  A variety of cooperative nuclear-atomic processes takes place instead: from internal conversion (IC) in neutral atoms to discrete, or resonance conversion  and electronic bridges in singly and doubly charged ions. It is didactic that in Ref. \cite{irwin}  the authors allegedly reported observation of the nuclear radiation from deexcitation of the isomeric state. Of course, that was  impossible, even with the isomer energy of 3.5 eV, as it was commonly accepted at that time. At this energy, deexcitation through the electronic bridges brought about by resonance conversion was the absolutely  dominating channel. The discrete conversion $R$ factor which plays the same role as internal conversion coefficient (ICC) in traditional
IC, comprises $R$ = 1600 \cite{antib,PL96,ka99}. Indeed, the effect of \cite{irwin}  was then attributed to the presence of extraneous $\alpha$ particles \cite{shaw,utt}.

   It was already noted in \cite{lars} that there was no observation of any IC signal for the singly charged ions Th$^+$, whereas quite a distinct signal from doubly charged Th$^{++}$ ions was detected. The absence of the Th$^+$ signal looked unusual. At the time this finding was tentatively attributed to the small amount of collected $^{229}$Th$^+$ ions. This fact was studied in more detail in  Ref. \cite{seif}. In that paper  the problem of a short isomeric lifetime in Th$^+$ was raised for the first time. In Refs. \cite{lars,seif}, recoil ions of $^{229}$Th from the $\alpha$ decay of $^{233}$U were collected in the stopping cell, and then transported through the nozzle, RPQ and mass separator to the final microchannel plate detector. The measurements were performed with doubly- and triply charged ions. As a result, certain limitations on the lifetimes in the singly- and doubly-charged ions were retrieved. Data \cite{lars,seif} comprised a topic for theoretical investigations \cite{volot, ka18}.
   The thorium puzzle received a further examination by  Seiferle \cite{seifdiser}, where further experimental details substantiating the problem can  be found.
   Data from \cite{lars,seif,seifdiser} became the starting point of appearance of the thorium puzzle,  actively discussed since then.

        We consider this question in more detail in the next sections. Various ways of possible solutions are pursued, such as the relation between the structural features of the electronic wavefunction in the ground state and the isomer lifetime,  a random resonance, and the most intriguing
reason --- a possible manifestation of the dependence on the ambient conditions. We show that fragmentation of the atomic wavefunctions is of great importance for the realization of each of the mechanisms. Note that it was already in papers \cite {TeZh,TeHars} that the decisive role of the electronic level fragmentation in the resonance state had been examined and proven soon after its discovery  in the case of RC in the 35.49 keV  $M1$ transition in $^{125}$Te$^{45+}$ ions \cite{atalah}. The resonance conversion  was discovered in this transition at the University of Bordeaux, thanks to a collaboration with the authors of the present investigation,  by means of a study of the dependence of the nuclear isomer lifetime on stripping the electron shell. The important role of  fragmentation was also considered in Ref. \cite{ka18}, assuming that the isomer energy is 7.6 eV. Herein we advance  and refine this approach, in the light of the latest values of the isomer energy as indicated above. We will see that the energy refinement is of crucial importance to solving the puzzle.

\section{The thorium puzzle and ways of its probable solution}

In Ref. \cite{seif,seifdiser}, along with neutral thorium atoms, restrictions on the isomer lifetimes in singly- and doubly charged atoms were obtained.     The halflife in any charged state of $^{229}$Th but singly charged one turned out to be  longer than one minute, as this was the maximum survival time of thorium ions under the achievable experimental conditions. On top of that comes the non-observation of an IC signal in singly charged Th$^+$ ions. Under experimental conditions of Refs. \cite{seif,seifdiser}, the
buffer-gas stopping cell was less than 10 ms. So the non-observation of the IC signal was attributed in \cite{seif} as pointing out to a potential reduced halflife of Th$^+$ ions. The latter was thus restricted by the experimental extraction time to 0.01 s as the halflife limit
$T_{1/2}^+< 0.01$ s given for Th$^{++}$; Th$^{+++}$ for more than one minute. This gave basis for separate speculations that the isomer energy might be more than 11.9 eV, which opened the IC channel. But the latter measurements mentioned above  make this supposition unlikely.
This  makes the case of singly charged ions most intriguing. Calculation \cite{ka18} results in a lifetime of not less than $\sim$1 s with a resonance conversion factor of $R\approx$ 5900. A question arises how one can explain the difference of two orders of magnitude. This situation got the name of thorium puzzle (\cite{seifdiser}).

   This situation has an explanation based on a physical ground. But before proceeding with detailed consideration, we note that the thorium puzzle arose even before the publication of Ref. \cite{ka18}, in attempts  to extract the $T_{1/2}^+$  from Ref. \cite{porsev}. Moreover, the discrepancy obtained appeared to be even two orders of magnitude greater than that indicated above. We remind that in Ref. \cite{porsev}, the authors calculated the probability of excitation of the nuclear isomer by means of a two-photon absorption through an electron shell in singly charged thorium ions, see Fig. 1a. The valence electronic configuration in the ground state was assumed to be $7s\ (6d)^2$. The first of the absorbed photons promotes an electron from the $6d$-  to the $7p$ state, transferring the atom as a whole to the excited level $(7s\ 6d\ 7p)_{J=5/2}$ with an energy of 3.084 eV. Then the same electron absorbs the second photon, appearing in the second intermediate state, from where it transfers to the nucleus already the total energy of both photons, itself returning to the ground state $6d$. The disadvantages of this scheme were considered in \cite{ka15,prc17}, and herein we will no longer dwell on them. Note, however,  that, firstly, the second intermediate state was not specified in Ref. \cite{porsev}; let us call it $x$.  In fact, in a situation close to a resonant one, quantum numbers of the nearest  $8s$ or $7d$ levels can be assigned to it. Then it is easy to see that this scheme excludes the most efficient resonant electron transition $8s - 7s$ for pumping the nucleus \cite{npa99}. This underestimates the efficiency of the entire circuit by up to three orders of magnitude: cf. the above work \cite{irwin},  where it was not taken into account that there is a decay through the $7s-8s$ electronic channel, with the discrete conversion   factor $R\approx$ 1600 expected to be valid instead.

   Secondly, only a part of the transition amplitude was actually calculated, starting with the second photon, namely, that of the $7p-x-6d$ circuit, with the atom remaining in the ground state in the end. This is illustrated in Fig. 1b. It may be  due to this error, that the calculated amplitude can be used in order to calculate the lifetime of the isomer in singly charged ions --- in view of the $T$-invariance of quantum electrodynamics, as shown in Fig. 1c. This erroneous judgment gave rise to a thorium riddle. The mistake is that the electron bridge in Fig. 1c is initiated by the resonance conversion, throwing up the $6d$ electron to the excited $x$ state ($8s$ or $7d$), from where it then transfers to the $7p$ state. This mechanism, however, does not represent the isomer lifetime, for much stronger conversion transitions are presented by the $7s - 8s - 7p$ electronic circuits \cite{antib}, and these transitions essentially remain outside the scope. As a result, our $R$ factor approximately equals the
coefficient $G$ defined in Ref. \cite{porsev}. Nevertheless, it is  the $\beta$ coefficient  of Ref. \cite{porsev}, which is analogue of the $R$ factor. But the latter is approximately twenty times less:
\be
R\approx\beta\approx\frac1{20}G_2\,.
\ee
This leads to the fact that, at an isomer energy of 7.8 eV, the divergence of paper \cite{porsev}  with the experiment turns out to be at least 20  times stronger than the one given above. Thus, a consistent calculation as in \cite{ka18} significantly reduced the severity of the thorium puzzle, though did not completely solve it yet. There was still a gap with the experiment of two orders of magnitude.

   \section{Dependence of the nuclear lifetime on the ambient conditions as an argument in solving the thorium puzzle}
   The next point is that  nuclei, decaying {\it via} resonance conversion, turn out to be vulnerable to the ambient conditions. Indeed, compare the case with
   traditional IC, when a bound electron, having received energy from the nucleus, leaves the atom. In contrast, in the case of resonance conversion, the conversion electron remains in the atom, sharing the obtained energy of the nuclear transition with the other electrons. In this case, the further electron fate evidently depends on the impact of the environment. First of all, this is collisional broadening of the electron state. Let us examine the question in more detail.

   In fact, the first 0.01 s fresh $^{229m}$Th$^+$ ions occurring from the alpha decay of $^{233}$U spend in a funnel in a buffer gas atmosphere at a pressure of about 40 mbar. Broadening from the collisions with the buffer gas molecules is proportional to the concentration of the gas, as well as the pressure. Bearing in mind that the collisional broadening  achieves two orders of magnitude at normal conditions, one can estimate that at the pressure of 30 to 40 mbar the broadening may be as large as of a factor of 3.
   Hence, in this way, from the initial four orders of magnitude, the paradox essentially decreases to an order of magnitude. In turn, this remaining discrepancy can be considered as a result of random resonance coincidence of the isomer energy with one of the intermediate electronic states \cite{ka06}. Moreover, as analysis shows, such an enhancement of the transition probability by an order of magnitude is not an unlikely event at all.

   The isomer undergoes deexcitation  through a great deal of the electronic bridge transitions like $7s - 8s - 7p$, with  a number of intermediate $8s$ and final $7p$ states, brought about by the interaction of the electronic configurations.
In these circuits, the first step --- lifting up the $7s$ electron to the $8s$ position --- is nothing but the bound, or resonance conversion (Fig. 2a). Because of the defect of resonance $\Delta$, this state cannot be the final state of the transition.   Emitting  a photon, the atom makes the final transition to one of the $7p$ states, restoring the energy conservation.
At a nuclear transition energy $\omega_n$, the $R$ factor is hence comprised by the sum of contributions of hunderds of intermediate states:
\be
R(\tau  L;\omega_n) = \sum_l \frac{\alpha_d^{(l)}(\tau  L;\omega_n) \Gamma_l/2\pi} {\Delta_l^2+(\Gamma_l/2)^2}
\ \equiv\ \sum_l R_l \,.    \label{e4}
\ee
In (\ref{e4}), $\alpha_d^{(l)}(\tau L)$ is the analogue of the ICC,  $\tau L$ indicate the type and  multipole order of the nuclear transition. The $\alpha_d^{(l)}(\tau L)$ values are practically independent of  the nuclear model, similar to the ICC.
However, they acquire the dimension of an energy in the case of bound (resonance) conversion. Therefore, they lose the sense of ICC, but they regain another meaning of residues in the sum  (\ref{e4}) \cite{ka18}. We return to this issue in section \ref{fragmen}. The $\alpha_d^{(l)}$ values are given by the squares of the conversion matrix elements between the initial $7s$ and final $8s$ electronic wavefunctions (e.g. \cite{atalah}).
$\Gamma_l$ is the full width of the intermediate state $l$, involving the atomic and nuclear widths, and $\Delta_l=\omega_n-\epsilon_l$ is the defect of the resonance
between the electronic and nuclear transitions ({\it e.g.}, \cite{ka18,atalah}). The $\Gamma_l$ values are typically around $10^{-8}-10^{-7}$ eV. For this reason, for a randomly chosen $\omega_n$, as a rule,  one meets the condition $\Delta_l \gg \Gamma_l$. The addition of $\Gamma_l$ in the denominator in (\ref{e4}) is important just near  resonances $l$, where the $R$ factor differs considerably from its regular value.

     As stated above, in the case of resonance conversion the $R$ factors from Eq. (\ref{e4}) play the same role as ICC --- in  traditional IC. That is, the probability of the isomer deexcitation per unit of time through the electronic bridge in Fig. 2a, induced by resonance conversion, is
\be
\Gamma_c (\tau L) = R\Gamma_\gamma (\tau L)\,,
\ee
where $\Gamma_\gamma$ is the  proper isomeric radiative width.

     Herein we assume that the expected energy of the isomer takes a random value within the range of $\Delta \omega_n = 0.4$ eV around 8.2 eV, $\omega_n \in$ [8.0, 8.4] eV. This energy interval is in agreement with the measurements of the isomer energy, cited previously. And we are thus interested in the probability $P(R_c)$ such that the $R$ factor must be at least $R_c$ or more, where $R_c$ may considerably exceed the regular value on the interval due  to a near coincidence with an intermediate electronic level $l$.
This imposes a condition on the $\Delta_l$ value.  Provided that
$R \geq R_c, R_c \gg 1$, one finds by means of (\ref{e4}) that
\be
P(R_c) = 2\Delta_l / \Delta \omega_n =
\sqrt{\frac{2\alpha_d \Gamma}{\pi R_c}} / \Delta \omega_n \,.       \label{e5}    \ee
It immediately follows from Eq. (\ref{e4}), that an increase of the atomic decay width $\Gamma_l$ by a factor of $K$  causes the same increase of the $R$ factor and the related shortening of the nuclear lifetime also by the same factor of $K$ times.  This is on one hand.
On the other hand, in a situation when the isomer energy is not precisely known, the probability of a random enhancement of the $R$ factor due to  coincidence with an intermediate electronic level increases simultaneously, but more slowly.  According to (\ref{e5}), it increases proportionally to the square root: $\sqrt{K}$ times.

\section{Influence of fragmentation on the $R$ factor of resonance
conversion: qualitative consideration}    \label{fragmen}

    In our case, we are in a position that the transition energy is unknown on the level of the required accuracy. Let us assess a scatter of the $R$ values which can be expected in such a situation. Our method will be similar to that used for treating  the dynamic enhancement of the effect of spatial parity nonconservation in the compound nuclei due to  fragmentation of the nucleon wave functions  in Ref. \cite{ufn}.
   Consider a schematic model, similar to the picket-fence model \cite{BM}.

     Our calculations were performed within the framework of the multiconfiguration Dirac-Fock (MCDF) method, taking into account the interaction of the electronic configurations, as in Ref. \cite{ka18}.      Concerning the electron configuration of the ground state in Th$^+$, different information can be found in the literature. One possibility is
      \be
  I: \qquad    7s(6d_{3/2})^2 \quad J=3/2   \label{6d} \ee
 (e.g., Ref. \cite{handbook}).  The other is
\be
   II: \qquad   (7s)^26d_{3/2} \quad J=3/2  \label{7s}  \ee
(e.g., Ref. \cite{NIST})
Our MCDF calculation shows that the ground state is structured according to Eq. (\ref{6d}). Nevertheless, these two single-electron states are practically degenerate, so that
there is a strong admixture of the configuration (\ref{7s}) in the ground state.
Therefore, in the main approximation, bearing in mind configuration $I$, one can put down the ground-state wavefunction as follows:
\be
\psi_0\sim 7s(6d)^2_{J=3/2} \,.  \label{wfg}
\ee
Resonance conversion transition transfers the $7s$ electron to the intermediate $8s$ state. For simplicity, suppose that taking into account of the interelectronic interaction splits the  $8s$ state into two levels. Then their wavefunctions may be expressed as follows:
\be
\psi_{in}\sim \frac1{\sqrt{2}}\left(\alpha|8s(6d)^2\rangle
+\beta |7s8s6d\rangle\right)_{J=3/2} \,,   \label{eq5}
\ee
with $|\alpha|\approx |\beta|\approx 1$.
If one uses wavefunction $I$ in the ground state, then  the first term in Eq. (\ref{eq5})  gives the predominant contribution to the $7s\to 8s$ transition. Oppositely, if wavefunction $II$ is used in the ground state, then  the main contribution will be due to the second term in Eq. (\ref{eq5}). Anyway, the $\alpha_d$ value, which is proportional to the square of  the wavefunction $\psi_{in}$ in (\ref{eq5}), becomes now $\alpha_d'$, which is twice smaller for each level:
\be \alpha_d'\sim \frac{\alpha_d}{2}\,, \ee
 as only the first term on the right side of (\ref{eq5}) gives a significant contribution. In contrast, the width of the state (\ref{eq5}) changes weaker, as both terms contribute. Taking into account that now we have two levels in the interval, the probability (\ref{e5}) becomes approximately
\be
P' (R_c) = 2\sqrt{\frac{2\alpha_d'\Gamma}{\pi R_c}}/ \Delta\omega_n =
\sqrt{2}P(R_c)\,,   \label{2lvl}     \ee
that is by a factor of $\sqrt{2}$ higher. Correspondingly, if one assumes splitting into $N$ levels, an amplification of $P'(R_c)$ by $\sqrt{N}$ times will be expected.

     Furthermore, one can compare what is said above with the results obtained within the   conventional Dirac---Fock method with no account of the configuration mixing,  making use of the ground-state configuration of type $I$. In this case, there is one intermediate $8s$ state, together with  two final states $7p_{1/2}$ and $7p_{3/2}$.  The calculated $\alpha_d(M1;\omega_n)$ value with an isomer energy $\omega_n$ =  8.2 eV is $\alpha_d$  = 5.6$\times$10$^9$ eV,  the radiative width of the intermediate $8s$ state
$\Gamma$ = 1.32$\times$10$^{-7}$ eV,  the $7s-8s$ transition energy being of 6.04 eV. Thus, the resonance defect would be tremendous in comparison with the width, making the resonance interaction unlikely. The fragmentation spreads the resonance strength far from the primary non-fragmented line to the inter-electronic-shell area. Note that $\alpha_d$ have an important physical  meaning of the residues of $R(\omega_n)$,  considered as an analytical function in the lower half-plane of the physical sheet of the nuclear transition energy:
\be
-2\pi i \Res R(\omega_n) \raisebox{-0.3ex} [0ex][-1ex]{ $\bigl|_{\omega_n = E_a^{(l)}}$} =
\alpha_d^{(l)} (\omega_n)  \,.
\label{resid}
\ee
Fragmentation of the $8s$ level,  accounting for a mixture of configurations, leads to an increase in the number of poles and to their spread over the energy region, accompanied with an adequate   decrease in the values of the residues, while
the total value of the residues approximately holds:
\be
\sum_l\alpha_d^{(l)} (\omega_n) = \alpha_d (\omega_n)\,.  \ee

Let us define $R_\text{mean}$ as  the weighted average \cite{BM}:
\be
R_\text{mean}=\int \rho(E_n-\lambda) R(\lambda)\ d\lambda \,.
\label{mn}
\ee
Here $\rho(x)$ is the averaging function. It is peaked around $x=0$, and falls off for large $|x|$. It is convenient to chose it as follows:
\be
\rho(x)=\frac{\Delta/2\pi}{x^2+(\Delta/2)^2}\,, \label{weight}
\ee
with  $\Delta$ characterizing the energy interval around $\omega_n$ over which the averaging is taken, $\omega_n$ = 8.2 eV, and $\Delta$ = 0.2 eV in our case. The weight function is  normalized at unity:
\be
\int \rho(x) \ dx = 1\,.
\ee
Inserting (\ref{weight}) into (\ref{mn}), neglecting in  first approximation  the dependence of $\alpha_d^{(l)} (\tau  L;\omega_n) \Gamma_l$ on the nuclear energy, and making use of (\ref{resid}), one arrives at the resulting expression:
\bea
R_\text{mean}= \sum_l \frac{\alpha_d^{(l)} (\tau  L;\omega_n) (\Gamma_l+\Delta)/2\pi} {\Delta_l^2+[(\Gamma_l+\Delta)/2)]^2} \doteq \nonumber \\
\sum_l \frac{\alpha_d^{(l)} (\tau  L;\omega_n) \Delta/2\pi} {(\omega_l-\omega_n) ^2 +(\Delta/2)^2}\,. \label{Rmean}
\eea
The last equality is derived taking into account that $\Delta \ggg \Gamma_l$.

\section {Results of the calculation}

     Bearing in mind the proper halflife of the isomer of approximately 2 h, we conclude that either $R_n = 2.4\times 10^5$  or $R_a = 7.2\times 10^5$  is needed for agreement with the experimental halflife of 0.01 s, depending on whether the influence of the ambient conditions is taken or not taken into account.
Passing to the calculation results, in Fig. 3a we present the $R$ values as calculated against the isomer energy $\omega_n$, based on the ground state $I$.
The $R$ values vary drastically on the interval, within well ten orders of magnitude in the peaks.
     This is very different from the interval around 7.6 eV considered in Ref. \cite{ka18}. To a large degree, the behavior of the plot $R(\omega_n)$ is due to the presence of the strong transitions into the intermediate states 8.148 eV, $J$ = 3/2
($\alpha_d^{(l)} = 2.39\times 10^9$,   $R_l=3.75\times 10^5$), and 8.211 eV, $J$ = 3/2 ($\alpha_d^{(l)} = 3.33   \times 10^7$,   $R_l = 3.15\times 10^4$).  The first state consists of the $8s$ component by 82\%. The second has a more miscellaneous composition. For the energy of $\omega_n$ = 8.2 eV the resulting $R$ value of $R=412066$ is obtained. This turns out to be nearly enough:
$R\gtrsim R_a$, in order to explain the observed lifetime. However, for the other possible values of the isomer energy on the interval, the $R$ factor  can be either much more or less, down to
$R_\text{min}= 7.24\times 10^4$ and $2.11\times 10^4$  for the isomer energies of $\omega_n$ = 8.02 and 8.4 eV, respectively, on the wings of the interval. In the former case, the $R$ value turns out to be in agreement with the data. This should  not  be taken as an argument in favor of the fact that the isomer energy is rather 8.2 eV or so. As it is said previously, experimental lifetime can be also explained as far as $R>R_n$, if a dependence on the ambient conditions is taken into account.

    The calculations were also performed with the ground state (\ref{7s}), corresponding to the configuration  $II$. The resulting plot of the $R$ factor against $\omega_n$ is presented in Fig. 3b. These values agree worse with experiment on the interval under consideration. Thus, $R$ = 67859 for $\omega_n$ = 8.2 eV, that is six times less. The peak created by the strongest transitions is shifted towards smaller values of the transition energies.

     Continuing our analysis, we note, that the mean $R_\text{mean}$ value on the interval   as obtained by means of (\ref{Rmean}) turns out to be $R_{\text{mean}} = 4.6\times 10^8$ and   1.6$\times 10^9$ for the cases $I$ and $II$, respectively.   At first glance, such big values may conflict with the other values  of  $R$ = 412066 or 67859, calculated above at $\omega_n$ = 8.2 eV.
However, this example teaches once again how the average values may not be indicative, while they are mainly achieved within the very
narrow vicinities of extremely sharp resonances. Actual values are greatly scattered around the mean value, between minimum $R_\text{min}$ = 10197 and maximal one $R_\text{max} = 1.59\times 10^{14}$ at $\omega_n$ = 8.48  and 8.14 eV, respectively.
For comparison, the minimal non-fragmented value is much less: $R_\text{min}^{(nf)}$ = 21. Correspondingly, the average non-fragmented value is also much less: $R_\text{mean}^{(nf)}$ = 25. Hence, there could be no question of agreement between theory and experiment in the absence of fragmentation.

     Now one can determine  the probability $P(R_c)$    that the $R$ factor will be $R>R_c$ under condition of random distribution of the isomer energy over the chosen interval $\Delta \omega_n = 0.4$ eV around 8.2 eV as follows:
\be
P(R)\raisebox{-0.3em}[0em][-1em] {$\bigl|_{R>R_c}$}=\frac1{\Delta \omega_n}\int\limits_{(\Delta \omega_n)} H\left(R(\omega)-R_c\right)\ d\omega  \,,    \label{Rc}
\ee
where  $H(x)$  is  the Heaviside step function:
\bea
H(x) \, = \,   \left\{
            \begin{array}{l@{\hspace{2cm}}l}
                 1  &   \mbox{for } x \geq 0  \\
                 0    &  \mbox{for } x < 0
            \end{array}
   \right.
\eea
    Integration in Eq. (\ref{Rc}) is performed over the assumed domain of the nuclear energy $\Delta\omega_n$.
At the minimum values, $P(R_{min})$ = 100\%.

     In Fig.  4 we present a plot of the calculated $P(R_c)$ values for both initial electronic configurations.  In the case of the ground configuration $I$,  the corresponding probabilities comprise 41 or 24 percent for $R_c=2.4\times 10^5$ or
$R_c  = 7.2\times 10^5$, respectively, if the influence of the ambient conditions is taken into account or is not. Thus, the calculated probability of a random enhancement of the $R$ factor by an order of magnitude, needed for agreement between theory and experiment, attains 41 percent, that is comparable with unity. If the influence of the ambient conditions were neglected, the probability of the agreement remained great enough: 24 percent, which remains to be quite a significant quantity, comparable with unity. For comparison, in the case of the ground configuration $II$, corresponding probabilities would comprise 25  and 14 percent, respectively.

    At the same time, one should not forget about consequences of fragmentation in the ground state. If it is fragmented into $N_g$ states, the weight  of the $7s$ state is also disseminated over them. Therefore, the interaction force is also diminished by $N_g$ times.
Partly, it is this effect which is responsible for diminution of the mean and minimal values of the $R$ factor.

\section{Conclusion}

The above analysis shows that in fact the experimental upper bound of the isomeric lifetime in the singly charged ions is at least two orders of magnitude smaller than regular theoretical predictions \cite{ka18}. This gave rise to the thorium puzzle. A possible solution that the isomer energy might be  more than 11.9 eV is not supported by the latest measurements.

    Another possible solution assumes a random coincidence of the isomer and one of the electronic levels. Calculations \cite{ka18} and the above analysis
demonstrate that such a coincidence can be expected with the probability of 24 to 41 percent. This is not a small  probability. One can say that it may reconcile  theory with the experiment.
A significant role in the formation of such a probability belongs to the fragmentation of the electronic levels, caused by  the configuration mixing. First of all,  the fragmentation
leads to spreading of the resonance strength over the energy domains between the electronic shells. This allows the resonance condition to be met with higher probability. Second, the fragmentation increases the  probability of fluctuations of the $R$-factor, as the qualitative consideration in Section \ref{fragmen} shows. The  picture is completely similar to the case of resonance conversion   in the 35.49-keV $^{125}$Te$^{45+}$ ions \cite{TeZh}. In that case, fragmentation led to the splitting of each of the electronic levels for the conversion electron essentially into two ones, which induced resonance conversion in full scale. For higher ions, the electronic level density decreased, rapidly ruling out the resonance conversion. A similar picture is observed in Th ions: ionization potential in Th$^{++}$ is 18.3 eV --- 6.2 eV higher than that in Th$^+$ \cite{NIST}. This  decreases the electronic level density, making the lifetime longer and giving rise to the thorium puzzle.

    Furthermore, an important role belongs to the collisional broadening of the intermediate levels formed in resonance conversion. It is remarkable that it brings about the dependence of the nuclear lifetime on the ambient conditions. This dependence manifests itself also in two respects.
    First, it may increase the observed value of the $R$ factor, up to two orders of magnitude at normal conditions, specifically,  by several times --- under the conditions of experiment \cite{lars}.
     Second, it further increases the probability of random fluctuations of the $R$ factor. As a result, allowance  for the collisional broadening increases the probability of a random coincidence, needed for agreement with experiment, to 41 percent.

     The dependence on the ambient conditions stands in the same row with the other effects, such as laser assistance of the isomer decay \cite{kabzon}, or mixing of the nuclear levels with different spins  \cite{zylic}, yielding acceleration of the isomer decay by  factors of few hundred. Each of these effects offer ways of manipulating  the nuclear processes, affecting the electron shell.

     It is worthy of dwelling on paper \cite{zylic}, where an extreme case of hydrogen-like ions has been considered. Broadening of the isomer state by a factor of seven hundred was predicted due to the interaction between the components of hyperfine structure, formed by vector coupling of the electronic angular momentum $j = s$ = 1/2 to the nuclear spins in the ground and isomeric states. Meanwhile, such an electronic configuration: $^1S_{1/2}$  in triply charged ions is considered as the most suitable for  creation of the nuclear-optical clock in the pioneering work \cite {peik}.  Estimates fairly similar to \cite{zylic} show that one should beware of broadening the isomeric line in this configuration by about 100 times. At the same time, a method of aligned states in triply charged thorium ions, based on the valence  $5f$ electrons \cite{camp}, is less susceptible to the broadening. This is because the $M1$ ICC for the $5f$ electrons are by three orders of magnitude smaller than those for the electrons of the $7s$ shell. For this reason, the  construction of a
nuclear-optical frequency standard on triply charged ions looks more promising. The dependence of the nuclear lifetime on the external conditions demonstrated above should not be overlooked.

\bigskip
\qquad
    The authors would like to express their gratitude to L. von der Wense for inducing discussions. They are also grateful to M. Pf{\"u}tzner  and E. Peik for helpful remarks.

%%%%%%%%%%  Fig 1
\newpage
\clearpage
\vspace{2cm}
%\begin{figure}
\begin{minipage}[t]{0.3\textwidth}
\includegraphics[width=\textwidth]{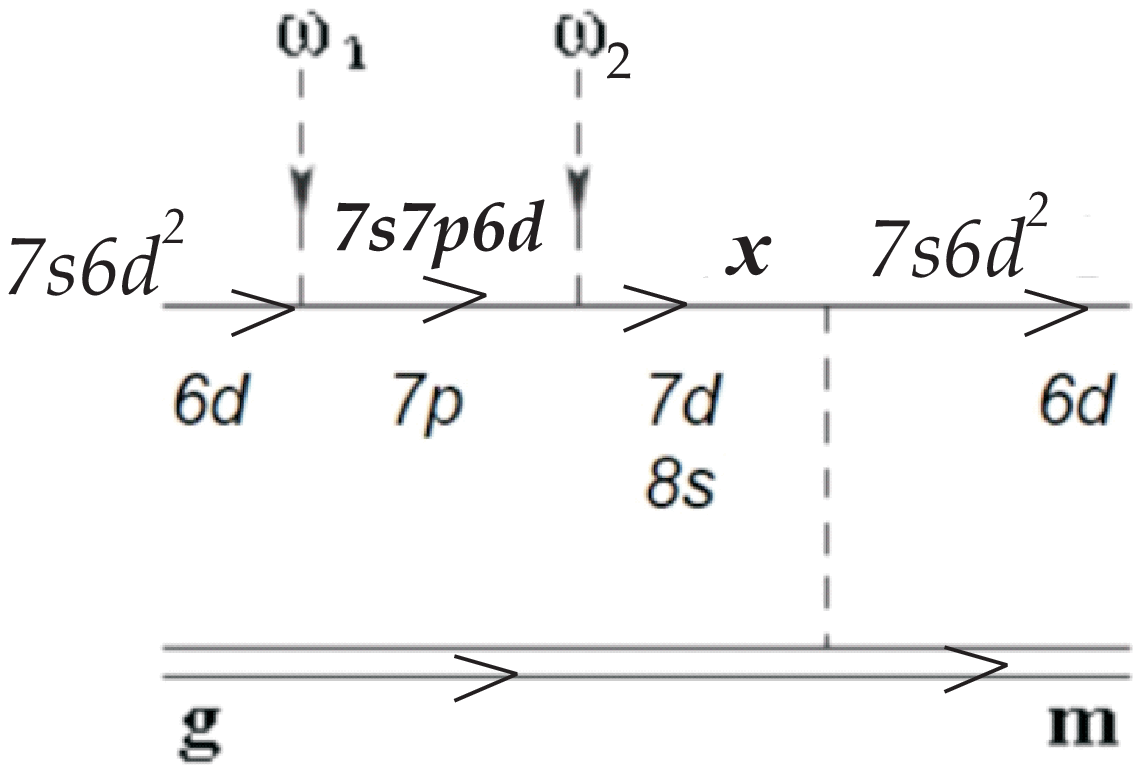}
\end{minipage}
%\vspace{-3.5cm}
\hspace{1em}\begin{minipage}[t]{0.3\textwidth}
\includegraphics[width=\textwidth]{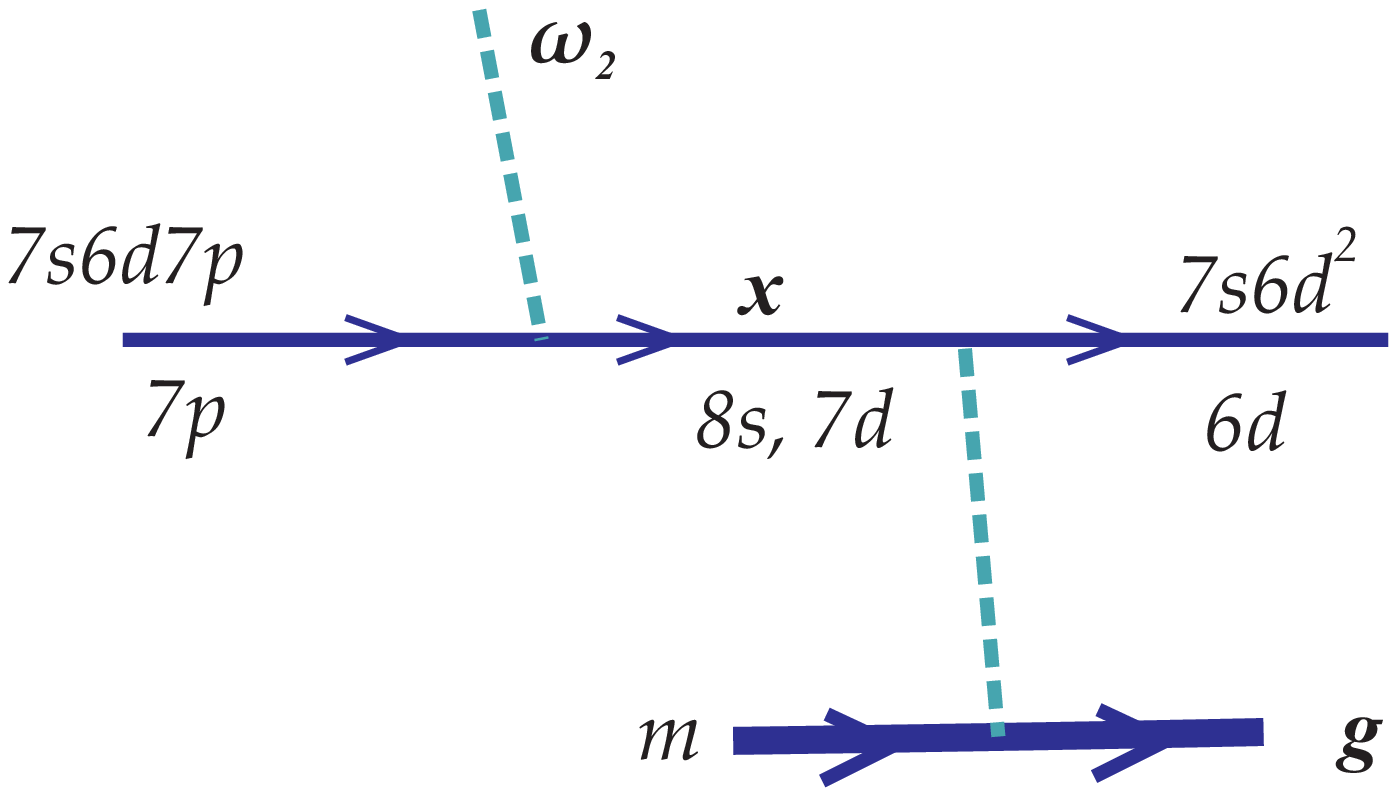}
\end{minipage}
\hspace{1em}
\begin{minipage}[t]{0.3\textwidth}
\includegraphics[width=\textwidth]{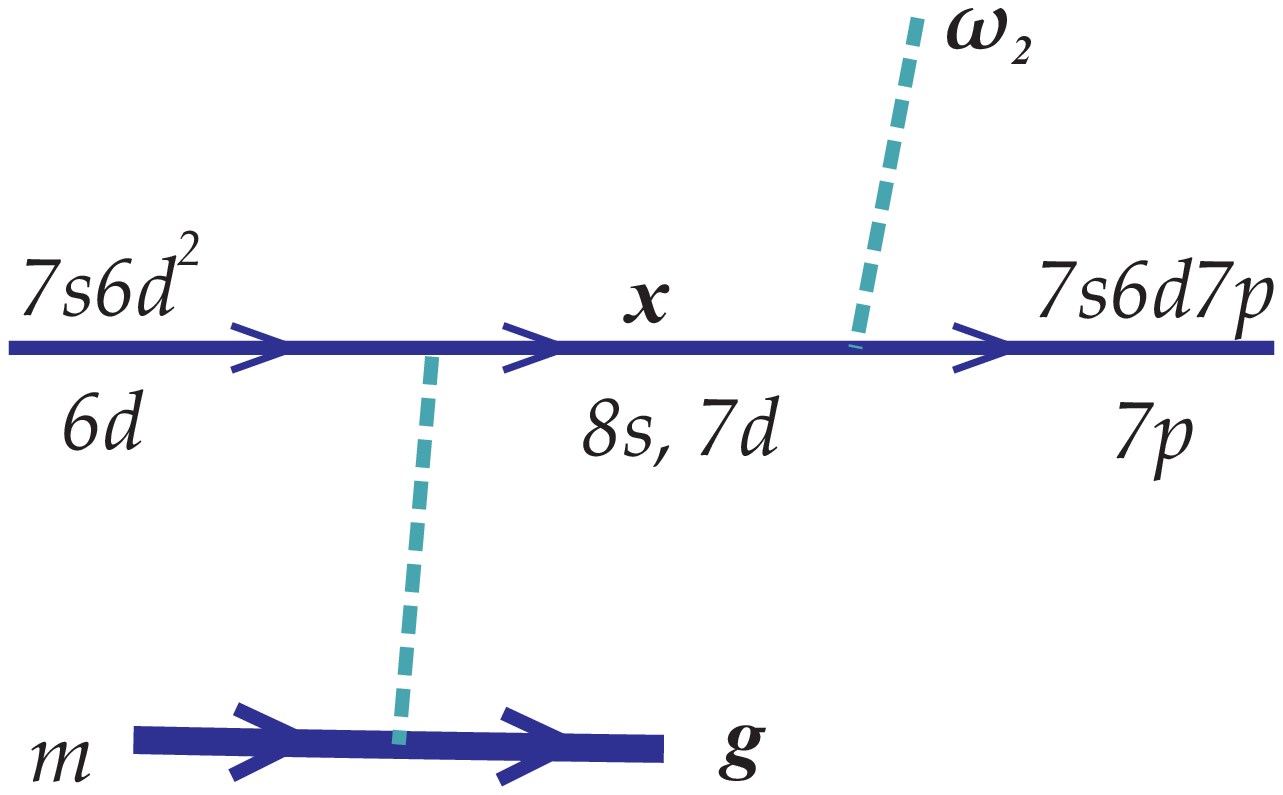}
\end{minipage}

%%\caption
\vspace*{1cm}{\small\bf Fig. 1.} \small Feynman graphs of: left:
two-photon scheme of isomer pumping \cite{porsev}; middle and
right: amplitude of the second photon absorption, calculated in
\cite{porsev}, and its reversal, which can be allegedly used for a
calculation of the isomer lifetime, respectively. The electronic
configurations together with a simplified classification of the
single-electronic states for pointing out the electronic circuits
are indicated.

%%%%%%%%%   Fig 2

\clearpage
\vspace{2cm}
\begin{minipage}[t]{0.45\textwidth}
\includegraphics[width=\textwidth]{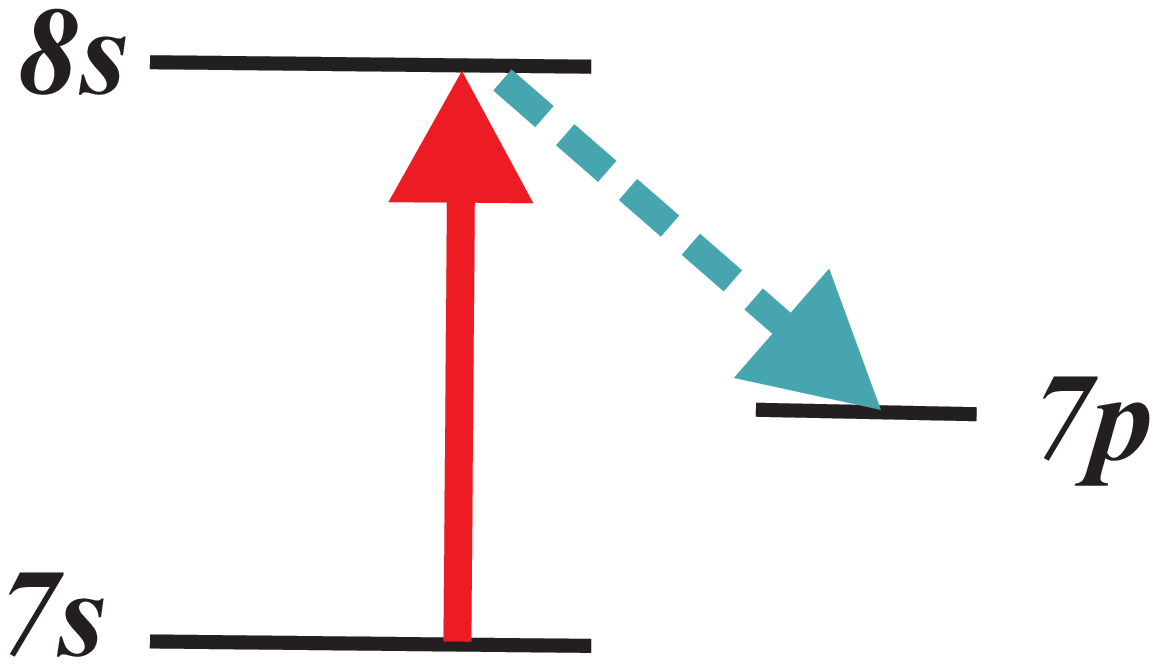}
\end{minipage}
\hspace*{5em}
\begin{minipage}[t]{0.45\textwidth}
\includegraphics[width=\textwidth]{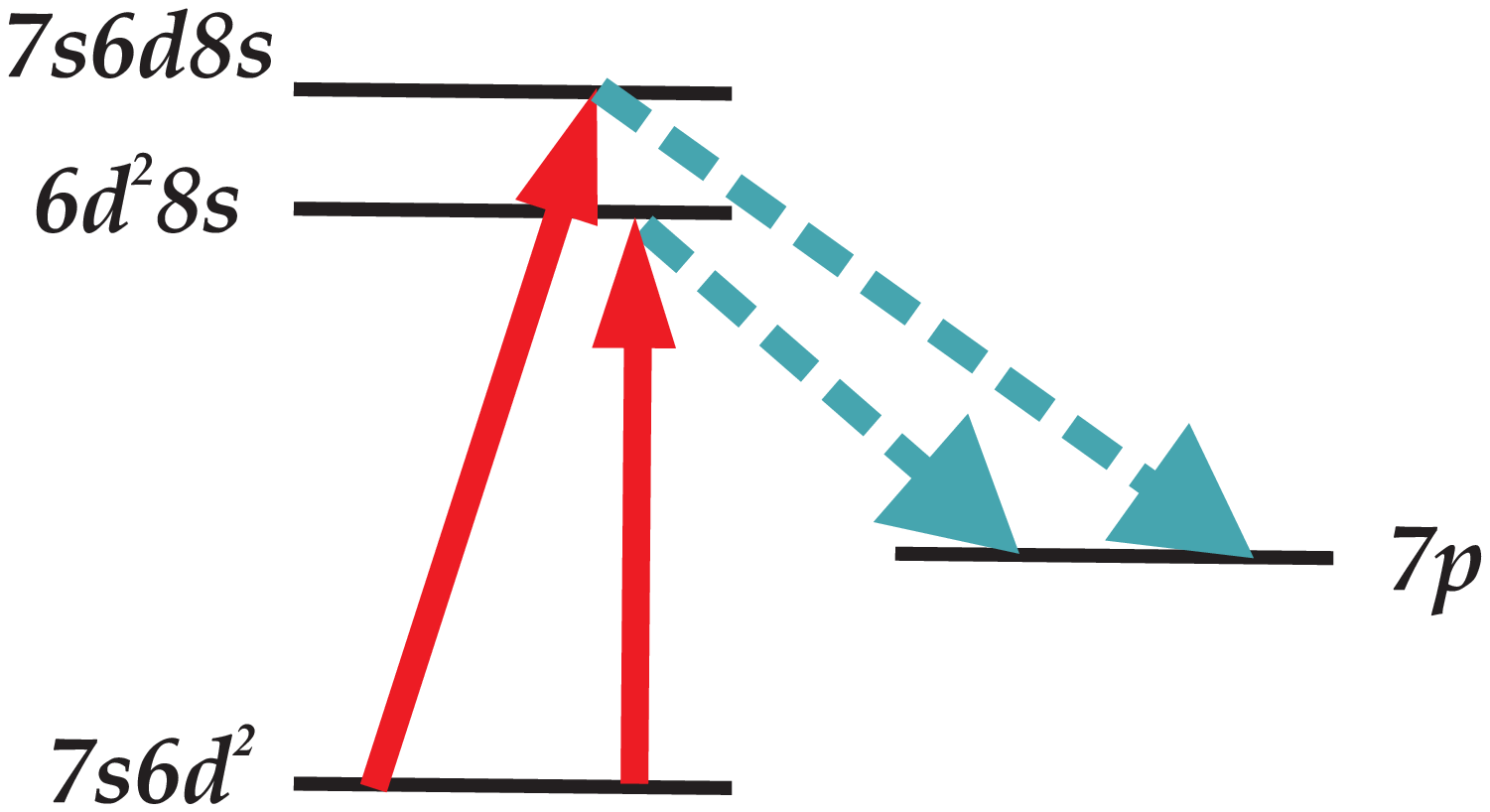}
\end{minipage}

\vspace*{2cm}
\vspace{1cm}{\small\bf Fig. 2.} \small Scheme of the resonance
electronic bridge deexcitation of the $^{229m}$Th isomer: left: as
anticipated in the absence of fragmentation; right: with
allowance for the fragmentation in the intermediate states.

%%%%%%%%%   Fig 3

\clearpage
\vspace{2cm}
\begin{minipage}[t]{0.45\textwidth}
\includegraphics[width=\textwidth]{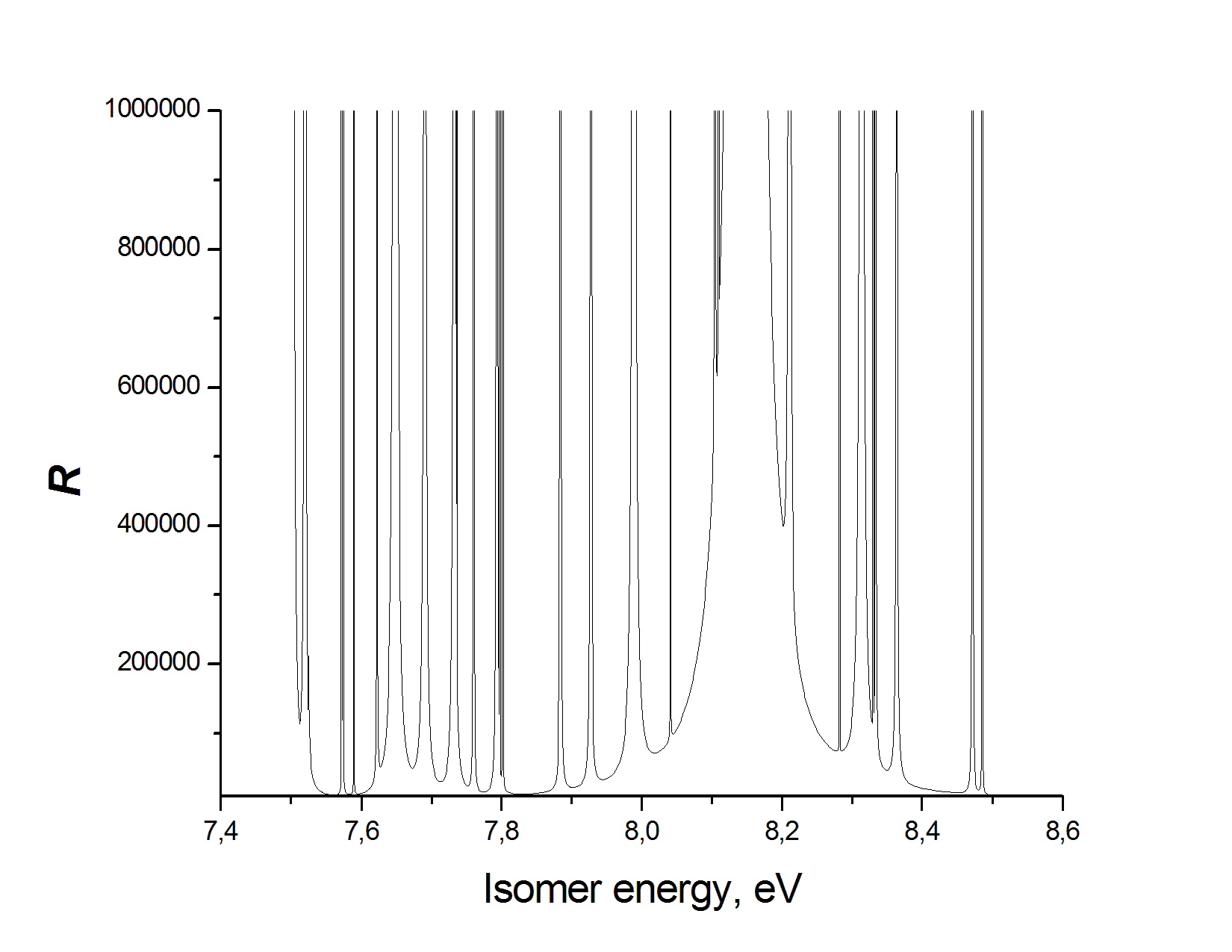}
\end{minipage}
\hspace*{5em}
\begin{minipage}[t]{0.45\textwidth}
\includegraphics[width=\textwidth]{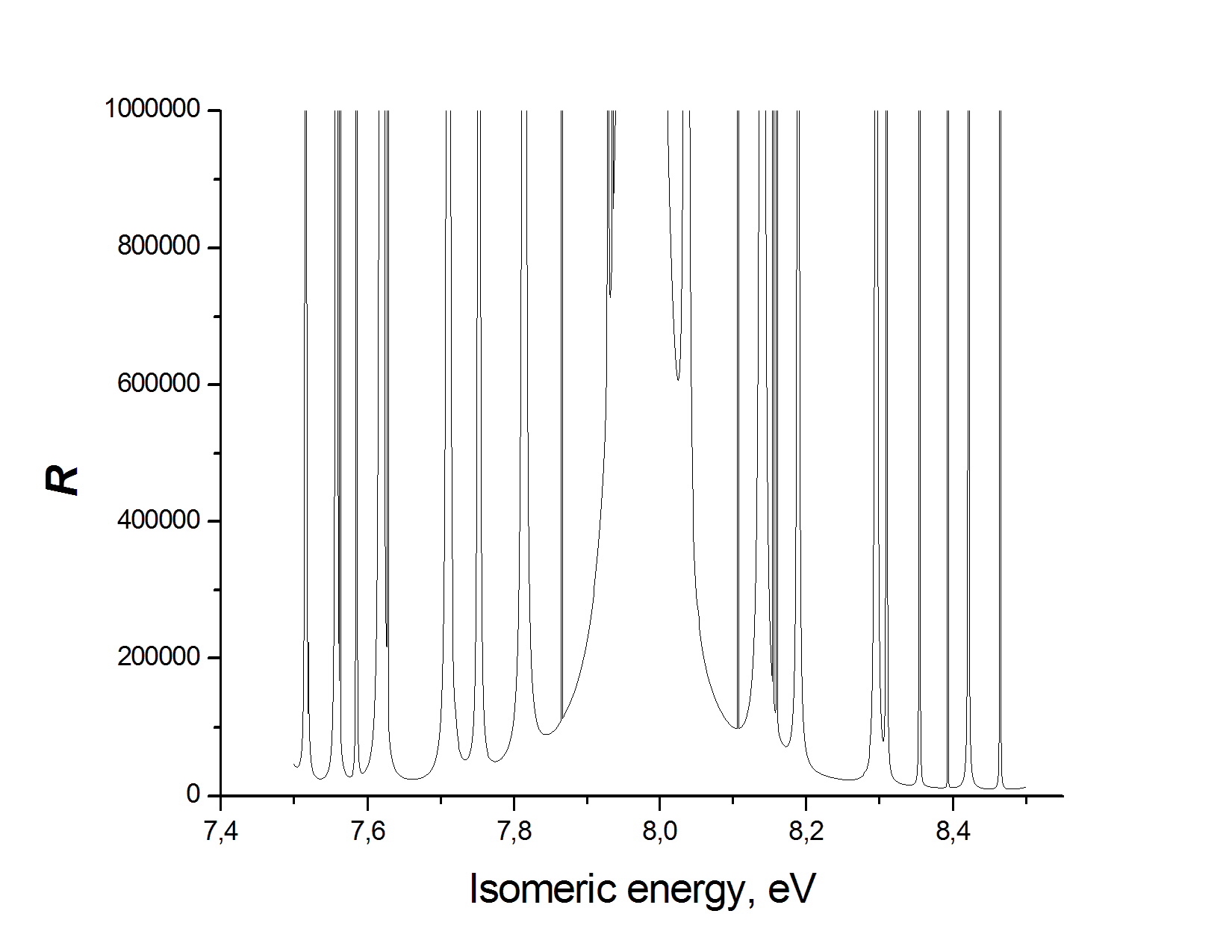}
\end{minipage}

\vspace*{2cm}
\vspace{1cm}{\small\bf Fig. 3.} {\small Calculated $R$ factor of resonance conversion  against the nuclear transition energy from the isomer to the ground state: left: calculation with the electronic ground-state
wavefunction $I$ (\ref{6d}); right: the same with the
wavefunction $II$ (\ref{7s}).}

\newpage
\bigskip
%%%%%%%%        Fig 4

\begin{minipage}[t]{\textwidth}
\includegraphics[width=\textwidth]{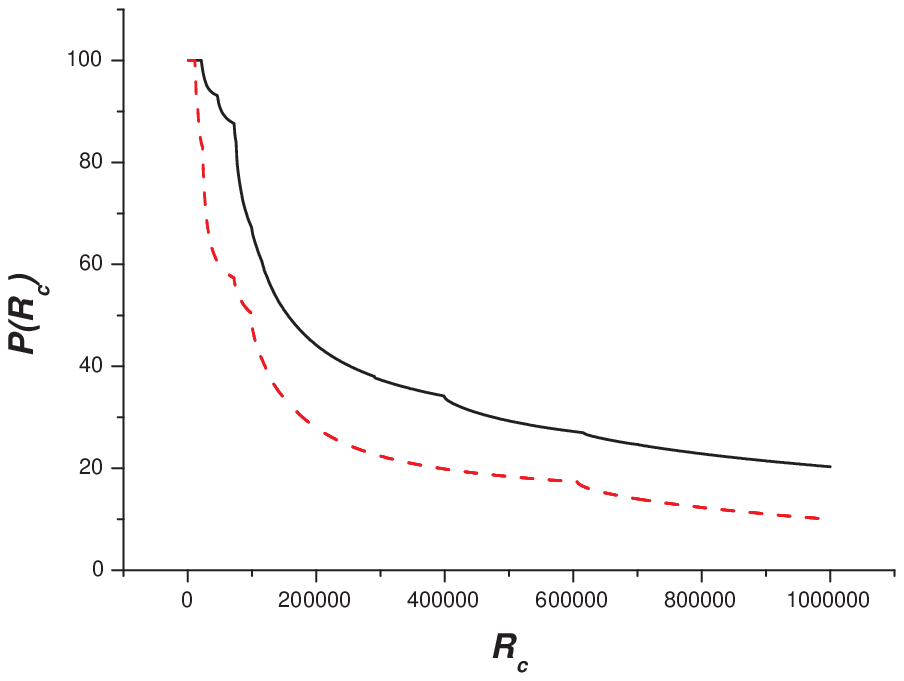}
\end{minipage}

\vspace*{2cm}
%\begin{center}
\vspace{1cm}{\small\bf Fig. 4. }\small
Probability $P(R_c)$, in percent,  that the $R$ factor will exceed $R_c$ under random choice of the isomer energy $\omega_n$ within the interval 8.0 -- 8.4 eV. Full (black) curve --- calculation with the ground-state wavefunction $I$ (\ref{6d}), broken (red) curve --- with the wavefunction $II$ (\ref{7s}).
%\end{center}

\end{document}